*Review*

# Membrane Trafficking in the Yeast *Saccharomyces cerevisiae* Model

**Serge Feyder, Johan-Owen De Craene, Séverine Bär, Dimitri L. Bertazzi and Sylvie Friant \***

Department of Molecular and Cellular Genetics, UMR7156, Université de Strasbourg and CNRS, 21 rue Descartes, Strasbourg 67084, France; E-Mails: sergefeyder@hotmail.com (S.F.); seahorse378@gmail.com (J.-O.D.C.); sbar@unistra.fr (S.B.); dimitri.bertazzi@gmail.com (D.L.B.)

**\*** Author to whom correspondence should be addressed; E-Mail: s.friant@unistra.fr; Tel.: +33-368-851-360.

**Abstract:** The yeast *Saccharomyces cerevisiae* is one of the best characterized eukaryotic models. The secretory pathway was the first trafficking pathway clearly understood mainly thanks to the work done in the laboratory of Randy Schekman in the 1980s. They have isolated yeast *sec* mutants unable to secrete an extracellular enzyme and these *SEC* genes were identified as encoding key effectors of the secretory machinery. For this work, the 2013 Nobel Prize in Physiology and Medicine has been awarded to Randy Schekman; the prize is shared with James Rothman and Thomas Südhof. Here, we present the different trafficking pathways of yeast *S. cerevisiae*. At the Golgi apparatus newly synthesized proteins are sorted between those transported to the plasma membrane (PM), or the external medium, via the exocytosis or secretory pathway (SEC), and those targeted to the vacuole either through endosomes (vacuolar protein sorting or VPS pathway) or directly (alkaline phosphatase or ALP pathway). Plasma membrane proteins can be internalized by endocytosis (END) and transported to endosomes where they are sorted between those targeted for vacuolar degradation and those redirected to the Golgi (recycling or RCY pathway). Studies in yeast *S. cerevisiae* allowed the identification of most of the known effectors, protein complexes, and trafficking pathways in eukaryotic cells, and most of them are conserved among eukaryotes.

**Keywords:** membrane trafficking; yeast; secretion; endocytosis; vacuolar protein sorting

## 1. The Yeast *Saccharomyces cerevisiae*

*1.1. Generalities and Historical View*

Yeasts are unicellular organisms, members of the Fungi kingdom, proliferating by budding or fission. They belong to the Dikarya subkingdom, which includes the Basidiomycota and Ascomycota phyla to which *Saccharomyces cerevisiae* belongs (Figure 1).

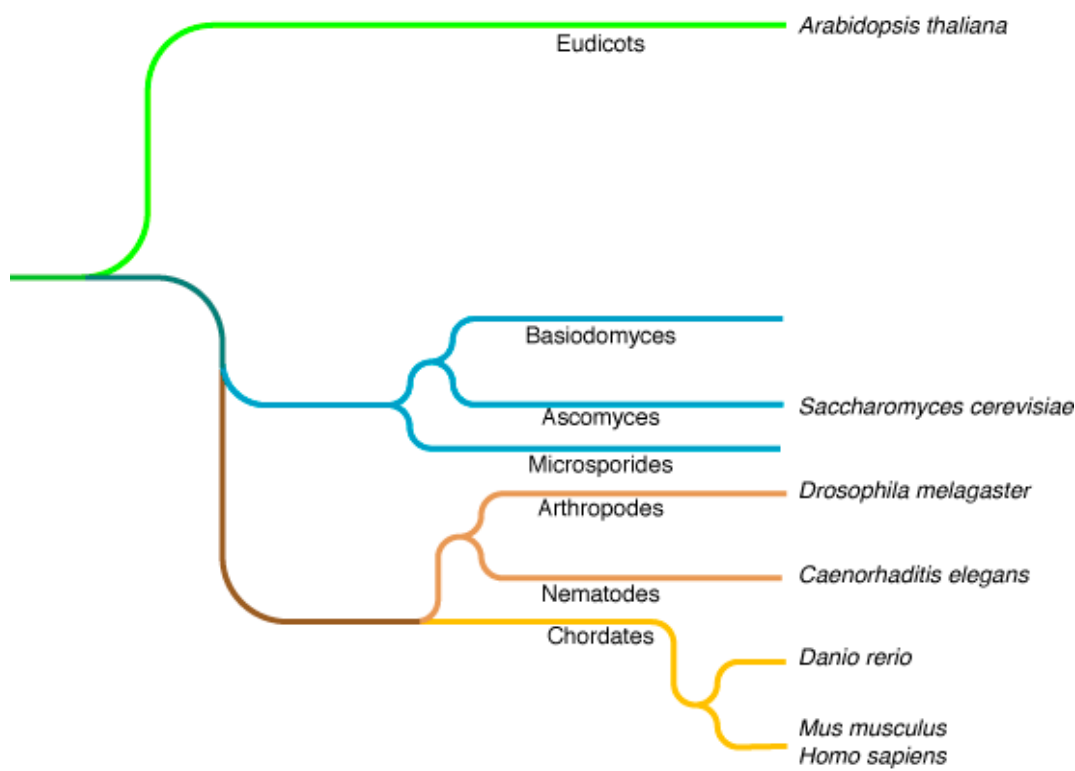

**Figure 1.** Phylogenetic tree of eukaryotes representing the distribution of cellular biology models.

The name "yeast" is a collective term, most often used to describe *S. cerevisiae*. The reason for this is historical, since the term yeast referred to a substance necessary for the process of fermentation during the production of bread and alcohol. The study of the fermentation process by yeast cells started in 1787 with Adamo Fabbroni (1748–1816) and his book, *Ragionamento sull'arte di far vino*. In 1857, Louis Pasteur discovered that the substance responsible for the fermentation process was a living organism. In the intervening time, baker's yeast was first identified, described and named in 1837 by the medical doctor and botanist Franz Meyen (1804–1840). Its scientific name, *Saccharomyces cerevisiae*, was given due to the high affinity for sugar (*Saccharo-*) of this fungi (*-myces*) and its use in brewing beer or "cervoise" (*cerevisiae*).

In 1875, Jacob Christian Jacobsen (1811–1887), founder of the Carlsberg brewing company, decided to set up a laboratory, "The Carlsberg Laboratory", to understand the physiology of yeast cells in order to better control the fermentation process. He hired Emil Christian Hansen (1842–1909), a renowned specialist in fungi and physiology. Hansen discovered that the fungi used by the industry were a mixture of different yeast species. He isolated and selected the clones that were the most interesting for fermentation, thus introducing the use of pure culture strains in industry [1]. He named

Unterhefe No. 1, the bottom-fermenting brewer's yeast used to produce the "Lager" since 1883; this strain is now termed *Saccharomyces carlsbergensis* [2]. Thanks to these pioneering studies and the development of experimental processes to isolate, maintain and culture pure yeast strains, *S. cerevisiae* has been playing an important role in research and has become a widely used eukaryotic model organism.

*1.2. Saccharomyces cerevisiae as a Model Organism*

*S. cerevisiae* is one of the most studied and best characterized models of eukaryotic organisms because it has several clear advantages.

Its first advantage is the shared complex intracellular organization with higher eukaryotes (Figure 2), as illustrated by its many fundamental contributions to the study of cellular processes such as cell signaling, membrane trafficking, lipid metabolism, and mitochondria import amongst others.

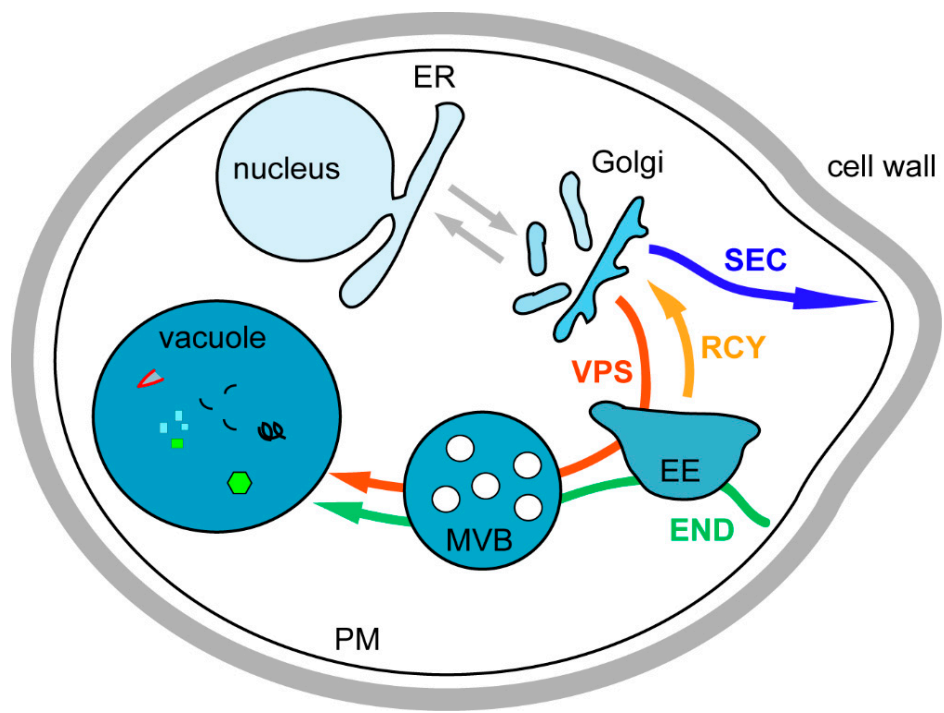

**Figure 2.** Representation of the different yeast intracellular trafficking pathways. Organellar soluble and membrane proteins are synthesized at the endoplasmic reticulum (ER) and transported to the unstacked Golgi (grey arrow). At the Golgi, these proteins are sorted into anterograde transport vesicles for ER resident proteins (grey arrow), into secretory (SEC) vesicles for plasma membrane (PM) and extracellular proteins (blue arrow), and into vacuolar protein sorting (VPS) vesicles for vacuolar proteins passing through endosomes (red arrow). The endocytic pathway (END) is used for internalization of PM proteins and extracellular medium components (green arrow). At the early endosomes (EE), proteins are sorted between those targeted for degradation into the vacuole, after maturation of the EE into the late endosome or multivesicular body (MVB), and those that are following the recycling pathway (RCY) to avoid degradation by being targeted to the Golgi (orange arrow).

Its second major advantage is its ease of handling compared to higher eukaryotes, as well as a short generation time of about 1.5 h in rich medium. Furthermore these cells can easily be stored short- or long-term on plates at 4 °C or in glycerol at −80 °C, respectively.

Its final and undoubtedly biggest advantage is its genetic tractability. Indeed, these cells can be stably grown as diploids or haploids; simple and fast techniques are available to change their genetic make-up [3]. The traditional technique is classic genetics by which haploid strains of opposite mating types (Mat a and alpha) can be conjugated resulting in a diploid which after sporulation (meiosis) can be dissected using yeast genetics to generate new haploid strains with combined genetic traits of the parental strains. With the rise of molecular biology, other genetic modifications could be introduced by the homologous recombination of exogenous DNA fragments with its genomic DNA to generate deletion mutants, fusion-proteins, or swap promoters [3].

Yeast *S. cerevisiae* research also benefited from being the first eukaryote to have its genome entirely sequenced [4]. Since it has been easy to generate mutants in yeast, its research community has devised a standard nomenclature for genes names, three letters followed by a number, usually given based on the biological function studied, as, for example, genes involved in uracil biosynthesis are *URAx*. In this nomenclature, gene names are always in italic and either in capital letters for the wild-type version (*URA3*), or in lower case letters for loss of function mutant versions (*URA3*), the corresponding protein is written with a capital letter followed by the two lower case letters and the number (*URA3*) (http://www.yeastgenome.org/help/community/nomenclature-conventions). A systematic gene naming was devised at the time of the complete genome sequencing of *S. cerevisiae* (in the 1990s) based on the gene's position on the chromosome. For example, for *URA3*, its nuclear open-reading frames (ORF) name is *YEL021W* where Y stands for yeast, E for the chromosome number (A for chromosome 1, B for 2), L indicates it is on the left chromosome arm, the number 021 its position on this chromosome arm counting from the centromere, and the last letter W for the strand on which it stands, either Watson (W, from 5' to 3') or Crick (C, the complementary strand).

Besides being able to modify the genome, genetic information can also be introduced on plasmids. These are characterized by their origin of replication, a 2 μ episomal origin based on a naturally occurring plasmid resulting in a high copy number plasmid (about 200 copies/cell) or a synthetic CEN/ARS centromeric autonomously replicating sequence with a low-copy number vector (1–3 copies/cell). These plasmids can carry one of a large set of either constitutively active or inducible/repressible promoters to modify gene expression. They also carry a selection marker which complements cellular auxotrophies (mutation in genes encoding amino acid or base metabolizing enzymes such as *LEU2*, *HIS3*, *TRP1*, *LYS2*, *URA3* or *ADE2*) and allows their maintenance into the cells after transformation, that is done by an easy to set up, low-cost and very efficient protocol [3,5]. All these tools make yeast a very attractive and amenable model system to study complex biological processes.

## 2. Introduction on Membrane Trafficking

In eukaryotic cells, membrane and soluble organellar proteins are generally translocated to the endoplasmic reticulum (ER) during synthesis and thus have to be transported to their correct target compartment in order to fulfill their function, undergo modifications, or be degraded. The cellular compartments of yeast *S. cerevisiae* and the different trafficking pathways connecting them are represented in Figure 2.

After being translocated to the ER, all newly synthesized proteins are directed to the Golgi apparatus where they are sorted between those transported to the plasma membrane (PM) or the external medium (exocytosis/secretory or SEC pathway) and those targeted to the vacuole either through endosomes (vacuolar protein sorting or VPS pathway) or directly (like the alkaline phosphatase or AP-3 pathway). Plasma membrane proteins can be internalized by endocytosis (END) and transported to endosomes where they are sorted between those targeted for vacuolar degradation and those redirected to the Golgi (recycling or RCY pathway) where they enter the secretory pathway to be readdressed to the PM. Most of the trafficking effectors, complexes and pathways described below were identified by studies in *S. cerevisiae*, and are very well conserved among eukaryotes [6–9].

## 3. Biosynthesis Pathway

### 3.1. Endoplasmic Reticulum to Golgi Transport

The biosynthesis pathway is a complex and highly regulated pathway allowing to correctly direct newly synthesized proteins to their target compartment [10,11]. The first step of this pathway consists in transporting newly synthesized proteins from the ER to the Golgi apparatus via the COPII-coated vesicles (Figure 3) [10]. The Sar1 GTPase is activated by its nucleotide exchange factor Sec12 localized at the ER membrane, this triggers the assembly of the prebudding cargo complex, composed of Sec23 and Sec24 subunits that sort cargos from ER resident proteins; this complex subsequently recruits the outer layer coat complex Sec13-Sec31, leading to membrane bending and COPII-coated vesicle formation [8,10,11]. At the ER, some transmembrane secretory cargos bearing sorting signals are directly recognized by multiple binding sites on the Sec24 subunit and concentrated into COPII-coated vesicles [12]. The COPII-dependent trafficking of soluble cargos and some transmembrane proteins depends on different ER sorting receptors [8]. Many secretory transmembrane proteins are concentrated into COPII vesicles by the Cornichon homologue Erv14, whereas the ER-export of type II membrane proteins (*N*-terminus in the cytoplasm) depends on the Erv26 membrane receptor [8,13,14]. The glycosyl phosphatidylinositol (GPI)-anchored proteins are sorted from plasma membrane proteins and incorporated into different COPII-coated vesicles via the p24 family members Emp24 protein (p24β) interacting with Erv25 (p24-δ) [15–17]. This shows that cargo sorting already occurs at the level of the ER [15]. The tetra-spanning membrane protein Erv29 acts as a cargo-receptor for ER sorting of different soluble luminal proteins [18]. Most of these cargo sorting mechanisms are conserved in mammalian cells [8]. Sorting and vesicle formation at the ER is a well-studied process involving specialized sites termed ER exit sites (ERES). Three ERES have been described in yeast, one being characterized by the soluble glycosylated α-factor, the second by the glucose transporter Hxt1 and the third by GPI-anchored proteins [17]. Moreover, the specific ER sorting of GPI-anchored protein that was discovered in yeast cells is conserved into polarized mammalian cells [9].

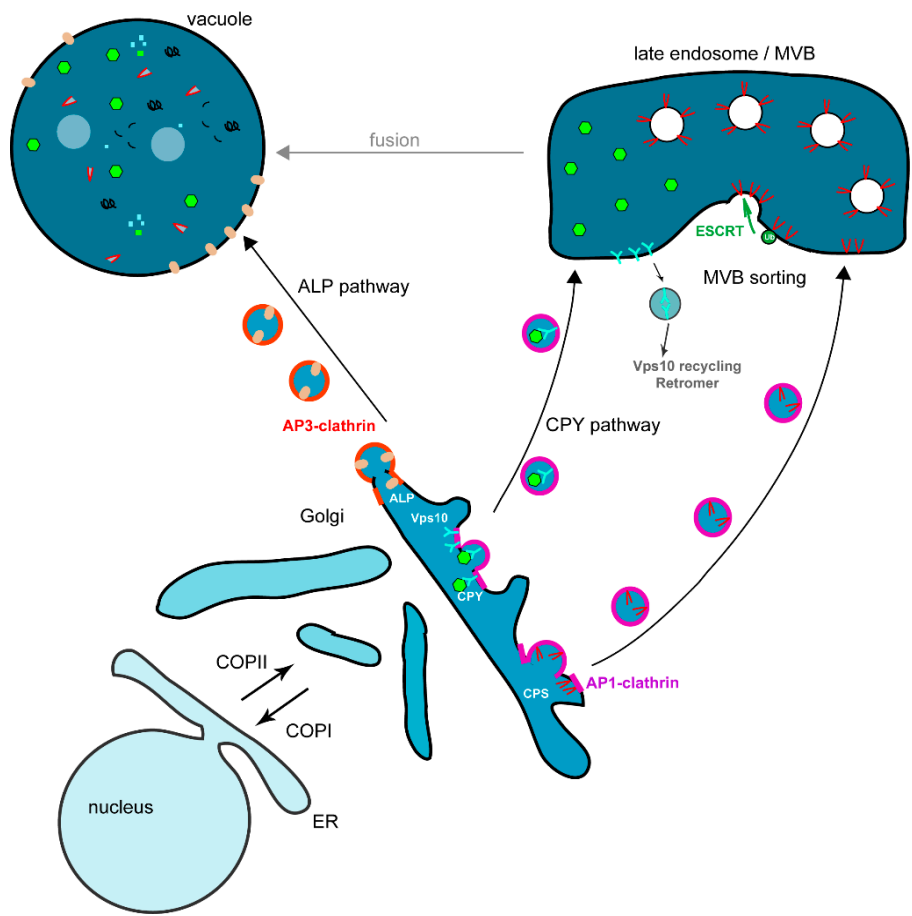

**Figure 3.** The Golgi to vacuole trafficking pathways. Two pathways transport proteins from the unstacked Golgi to the vacuole, either directly via the ALP pathway, or via the endosomes for the VPS CPY pathway. In the ALP pathway, the alkaline phosphatase ALP is packaged into vesicles through adaptor AP-3 interaction which in turn recruits clathrin. Vesicles are then fusing with the vacuole. The CPY pathway transports different vacuolar proteins, either membrane-bound as the CPS (carboxypeptidase S) or soluble as the CPY (carboxytpeptidase Y) pro-protease. The soluble CPY is recruited into the Golgi lumen through the binding of its receptor Vps10. Both CPY and CPS are loaded onto vesicles via the AP-1 adaptor and clathrin coat. After budding, vesicles are transported to the endosomes where they fuse liberating CPS to the endosomal membrane and CPY into the lumen. Vps10 is sorted away from the endosomal membrane to be recycled back to the Golgi for a new round of CPY transport by being loaded onto vesicles initiated by the Retromer complex. On the other hand CPS is loaded onto vesicles, which will bud into the endosomal lumen to give rise to the multivesicular body (MVB). This process requires first the ubiquitination of CPS as a sorting signal for the ESCRT (endosomal sorting complex required for transport) machinery (ESCRT-0 to -III), which will cluster CPS into invaginated endosomal membrane and pinch off vesicles. The MVB will then fuse with the vacuole and liberate its content into the vacuolar lumen to be processed.

These vesicles are transported to the Golgi where they are first tethered to the membrane by the multimeric guanine nucleotide-exchange factor TRAPP (transport protein particle) complex prior to specific Bet1 SNARE complex dependent-fusion which requires the GTPase Ypt1 [11]. COPI-coated vesicles mediate the retrograde transport from the Golgi to the ER for retrieval of essential components of the transport machinery, as well as ER resident proteins (Figure 3) [19]. The soluble ER resident proteins have a HDEL retrieval signal recognized by the Erd2 receptor [20], whereas ER-resident SNARE proteins and integral membrane proteins have a *C*-terminal KKXX retrieval motif recognized by the Rer1 receptor [21]. The Golgi to ER retrieval depends on the Dsl1 tethering complex that is composed of three subunits, Dsl1, Dsl3 and Tip20, and interacts with COPI coat and ER SNARE receptors [22,23]. The COPI and COPII coats as well as the tethering complexes and the GTPases are conserved from yeast to human [11]. The cargo proteins progressing from the *cis*-Golgi to the *trans*-Golgi network (TGN) can undergo maturation and modifications. In recent years there has been a controversy about the intra-Golgi trafficking of proteins. Indeed, until recently, the Golgi was considered as a "static" compartment, bearing the *cis*-, *medium*- and *trans*-Golgi saccules, with COPI-coated vesicles mediating the transport of materials along the Golgi. New *in vivo* studies performed in the laboratories of Ben Glick and Aki Nakano and mainly based on high-resolution live microscopy in yeast cells, have proposed an alternative model of cisternal maturation for intra-Golgi protein transport [24,25]. Indeed, cisternal maturation was visualized in yeast cells because Golgi cisternae are dispersed into the cytoplasm and thus resolvable using fluorescent microscopy. In this new model, the Golgi forms *de novo*, matures (with its cargoes), and finally dissipates with COPI coated vesicles mainly mediating retrograde trafficking of components that have to return to the ER [24,25]. The strong experimental evidences obtained in yeast and mammalian cells to explain how proteins are transported across the Golgi are recapitulated into a review written after a Golgi meeting gathering many researchers involved in the field [26]. At the TGN level, cargo proteins are sorted into secretory vesicles destined to fuse with the plasma membrane, while others are packaged into transport vesicles which fuse either with the endosomes (VPS pathway) or with the vacuole (AP-3 pathway) (Figure 3). Golgi proteins are retained in this compartment thanks to the presence of retention signals in their protein sequence [26].

*3.2. Transport from the Golgi to the Plasma Membrane*

Cargo proteins destined to the plasma membrane (PM) are loaded into vesicles at the TGN and follow the secretory (SEC) pathway (Figure 2). The secretory pathway was the first trafficking pathway clearly identified thanks to the work of George Palade showing vesicular transport from ER to Golgi and vesicular budding from the Golgi in pancreatic exocrine cells of the guinea pig [27]. The work on yeast cells mainly performed in Randy Schekman's laboratory in the early 1980s, allowed the identification of the molecular mechanisms by which this transport occurred. They have isolated yeast *sec* mutants unable to secrete the extracellular enzyme invertase and most of these *SEC* genes encode key effectors of the secretory machinery [28,29]. The 2013 Nobel Prize in Physiology and Medicine was awarded to Randy Schekman for this discovery; this Nobel Prize is shared with James Rothman for his mammalian cell-free assay recapitulating the ER to Golgi transport pathway and to Thomas Südhof for his discovery of the molecular machinery that regulates vesicular release at synapses. These SEC vesicles are then directed to the polarized sites of growth by tropomyosin-actin cables and delivered to the PM with which they are tethered by the exocyst complex (composed of Sec3, Sec5,

Sec10, Sec6, Sec8, Sec15, Exo70 and Exo84) prior to Snc1 SNARE complex dependent fusion in a Sec4 GTPase manner [30,31]. Studies suggest the existence of two different secretion pathways, since secretory vesicles can be separated in two groups by equilibrium centrifugations and have different protein compositions, and genetic studies show that one type goes directly from the TGN to the PM, while the second transits by an intermediate endosomal compartment [32,33]. In polarized mammalian cells, there are also two types of secretory vesicles, one targeted to the apical membrane and the other to the basolateral membrane [6]. Despite these differences, both cases require lipids for the secretory vesicle scission from the late Golgi apparatus. One is diacylglycerol (DAG), which favors scission by destabilizing the membrane due to its conical shape, and the other is phosphoinositide phosphatidylinositol 4-phosphate (PtdIns4*P*), which mediates the recruitment of specific effectors required for formation of the secretory vesicle [6].

*3.3. Transport from the Golgi to the Vacuole via the Endosomes, the VPS Pathway*

Most of the vacuolar targeted membrane proteins follow the vacuolar protein sorting VPS pathway (Figure 2). The effectors required for this pathway were identified by different sets of genetic screens, generating *vpt* and *vpl* mutants that were later renamed *vps* [34]. Based on their vacuolar morphology, these different VPS genes were sorted into class A to F [35]. Some Vps proteins belonging to the same class were later shown to be associated into protein complexes [34]. At the Golgi, newly synthesized vacuolar proteins are sorted through their interaction with different effectors such as the adaptor AP-1 complex [36], Golgi-associated gamma-adaptin ear homology domain Arf-binding proteins (GGA) [37–39] and epsins Ent3/Ent5 [40,41]. These different adaptors concentrate the cargos and ensure the recruitment of the clathrin coat to form the vesicles at the TGN membrane [34]. The vesicles fuse with the endosomal membrane where vacuolar soluble proteins (e.g., vacuolar proteases) are delivered to the lumen of the endosomes and vacuolar membrane proteins (e.g., the vacuolar ATPase) remain at the endosomal membrane. The endosomal tethering complex is termed CORVET (core class C vacuole/endosome tethering) and contains the class C Vps11, Vps16, Vps18 and Vps33 proteins and the class D Vps8 and Vps3 proteins [42]. In the VPS pathway, the final delivery of the sorted membrane, or soluble cargos, is mediated by fusion between the late endosome and the vacuole (Figure 3). This process requires the tethering HOPS (homotypic fusion and protein sorting) complex that is closely related to the CORVET complex and is composed of the class C core (Vps11, Vps16, Vps18 and Vps33) bound to the class B Vps39 and Vps41 proteins, the fusion depends also on the vacuolar SNARE Vam3 and the Ypt7 GTPase [42,43].

3.3.1. Vacuolar Targeting of Soluble Proteins

The best-studied soluble protein targeted to the vacuole through this VPS pathway is the carboxypeptidase Y CPY. Most of the Vps proteins were identified by complementation of their corresponding *vps* mutants isolated in mutagenesis screens having as read-out mis-secretion of soluble CPY protease in the extracellular medium instead of their correct delivery to the vacuole [34]. At the Golgi, soluble pro-enzyme CPY binds its transmembrane receptor Vps10 (homologous to the mammalian mannose-6-phosphate receptor M6PR) for transport. Vps10 and its bound cargo are specifically addressed from the TGN to the endosomes (Figure 3). The endosomal tethering CORVET complex interacts with the Rab Vps21 GTPase [42]. CPY is released in the endosomal lumen and reaches its final destination after fusion between the endosome and the vacuole. Vps10 recycles to the TGN via the retromer composed of two distinct subcomplexes, the Vps35-Vps29-Vps26 complex involved in cargo selection and the Vps5-Vps17 complex that has a structural role [34,44].

3.3.2. Vacuolar Targeting of Membrane Proteins via Multivesicular Body (MVB) Sorting

Newly synthesized membrane proteins that have to reach the vacuolar lumen require an additional sorting step at the endosome. To be delivered to the vacuolar lumen, these membrane proteins have to be sorted into vesicles budding into the lumen of the late endosome or multivesicular body (MVB). The MVB fuses with the vacuole and liberates its content into the vacuolar lumen. This inward budding depends on a specific protein machinery mainly composed of class E Vps proteins. The cargo membrane proteins are ubiquitinated by the Rsp5 ubiquitin ligase to ensure their recognition by the MVB sorting machinery (Figure 3) [45]. Ubiquitin is a highly conserved 76 amino acid peptide interacting with MVB effectors bearing ubiquitin interacting motifs (UIM). The MVB sorting machinery is composed of class E Vps proteins assembled into four different protein complexes termed ESCRT (endosomal sorting complex required for transport) (Figure 3) [34,46]. The ESCRT-0 complex composed of Vps27/Hrs and Hse1/STAM, is recruited at the MVB membrane via the specific binding of the FYVE domain of Vps27/Hrs to the endosomal phosphoinositide PtdIns3*P*. Vps27 has also an UIM motif to recruit the cargo and a motif to recruit Vps23/TSG101 a component of the ESCRT-I complex [47]. The ESCRT-I complex binds to the ubiquitinated cargos and is required for its MVB sorting, this was the first identified ESCRT complex [48]. ESCRT-II is physically linked to the endosomal membrane via PtdIns3*P* binding mediated by the GLUE domain of Vps36, to the ubiquitinated cargo via Vps36 and to the Vps28 subunit of the ESCRT-I complex, it also interacts via its Vps25 subunit with the Vps20 ESCRT-III subunit. ESCRT-III is involved in the vesicle scission via the Vps4 AAA-ATPase to generate force [49]. These different interactions between the cargo, the ESCRT-0, -I, -II and -III complexes allow the sorting of the cargo [46].

The most studied yeast biosynthetic cargos undergoing this MVB sorting are the carboxypeptidase S Cps1, the endopolyphosphatase Phm5 and the transmembrane protein Sna3 [50]. The MVB sorting of the ubiquitinated-cargo Sna3 is in some way different because it mainly depends on its direct binding to the ubiquitin ligase Rsp5 to be properly targeted to the vacuole [51].

*3.4. Transport from the Golgi to the Vacuole via the AP-3 Adaptor or ALP Pathway*

In the ALP (alkaline phosphatase) pathway vesicles formed at the TGN are directly targeted to the vacuole (Figure 3). The existence of this pathway has been shown by studies on the alkaline phosphatase ALP [52]. The ALP enzyme is correctly delivered to the vacuole even in *vps* mutants that are deficient for fusion of vesicles from the TGN with endosomes such as the *pep12/vps6* SNARE mutant. This indicates the existence of a pathway from the TGN to the vacuole, which is independent of endosomes and was termed the "ALP pathway" [52]. This pathway relies on the HOPS tethering complex and Ypt7 GTPase for vesicular fusion with the vacuole [43]. The machinery specifically required for the ALP pathway has been later identified and contains the adaptor complex AP-3 composed of different subunits Ap16, Ap15, Apm3 and Aps3 [53]. At the TGN, the cargos following the ALP pathway are specifically sorted into the AP-3 clathrin coated vesicles. One other important cargo using this pathway for its vacuolar membrane delivery is the transmembrane SNARE Vam3, which is required for fusion of the TGN vesicles with the vacuole [43].

## 4. Endocytosis

The endocytic pathway (END) allows eukaryotic cells to internalize plasma membrane proteins and extracellular medium and deliver them to the cell interior for further transport through early endosomes and late endosomes/MVB before reaching either the vacuole/lysosome for degradation or transport back to the TGN for recycling (Figure 2) [7]. Generally a distinction is made between two types of endocytosis: fluid-phase endocytosis during which endocytic vesicles form and sequester extracellular material non-specifically, considered constitutive; and receptor mediated endocytosis, considered inducible because it is triggered after the binding of a ligand by its receptor.

In *S. cerevisiae*, endocytosis was shown to be mainly actin dependent based on genetic screens allowing the isolation of *end* (endocytosis) mutants [54,55]. Indeed, characterization of these *end* mutants and others isolated in different genetic screens revealed that the mutated genes are coding for actin (Act1), actin related proteins (Arp2/Arp3), actin polymerization effectors WASP (Wiskott-Aldrich Syndrome Protein) (Las17) or actin cytoskeleton linked-effectors (Sla2, Pan1, Rvs161/167) [55–57]. The role of these different proteins into the endocytic internalization step is described in Figure 4.

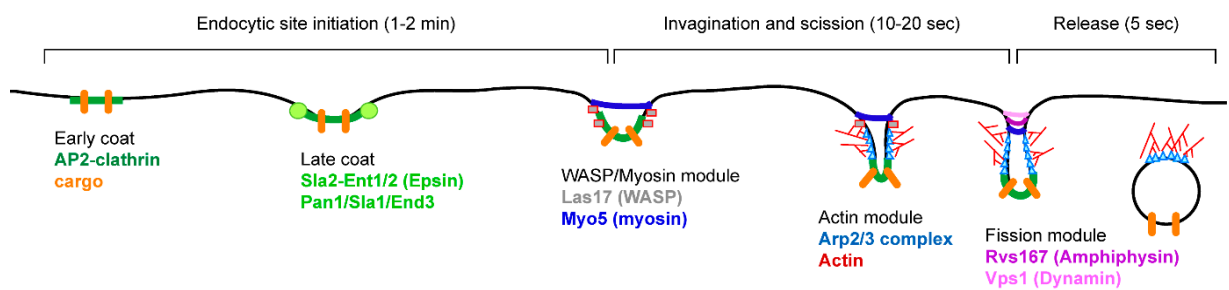

**Figure 4.** The endocytic internalization process. Endocytosis sites are first initiated by the recruitment of clathrin via the AP-2 adaptor complex (early coat) to cluster cargo proteins to the endocytic invagination sites at the plasma membrane (PM). Initiation is terminated by the recruitment of ANTH- (AP180 *N*-terminal homology) and ENTH- (epsin *N*-terminal homology) domain containing proteins Sla2 and epsins Ent1 and Ent2 to form the late coat. Invagination starts by the recruitment of the WASP/Myosin module to initiate actin branched polymerization by the actin module (Arp2/3 complex and actin). Myosin is squeezing the PM while the actin module expands the invagination. Once the invagination in long enough, the fission module starts pinching off vesicles that are released into the cytoplasm.

More recently, live-cell imaging studies in yeast have allowed the dynamics and functions of these actin-dependent effectors to be deciphered during endocytic internalization [58]. These studies revealed the succession of steps leading to the internalization (initiation, invagination, scission and vesicle release) and the different protein complexes required at these different steps (Figure 4). The current model for actin based endocytic internalization relies also on immuno-electron-microscopy studies on yeast cells showing the position of the different effectors along the endocytic invagination [59,60]. Dynamic actin structures are associated with the endocytic vesicles, and favor their formation, their release from the plasma membrane, and their transport into the cell cytoplasm. Initiation starts by clathrin binding to the endocytic site via the AP-2 adaptor proteins complex, and concomitantly, cargo proteins are recruited to form the early coat (Figure 4). Then the epsins Ent1/Ent2 proteins interact with the phosphoinositide PtdIns(4,5)$P_2$ via their ENTH (Epsin *N*-terminal homology) domain, and this allows the recruitment of the Pan1-Sla1-End3 complex (late coat). This late coat is crucial for recruiting the actin polymerization machinery (WASP/Myosin module), which will generate the invagination of the plasma membrane. Las17/WASP and type I myosins Myo3/Myo5 are activators of the Arp2/3 protein complex. The Arp2/3 complex nucleates the branched actin filaments to allow the formation of the endocytic vesicle (Actin module). Along the invagination structure, scission occurs requiring amphiphysins Rvs161/Rvs167 which sense membrane curvature via their BAR (Bin1/amphiphysin/Rvs167) domain and the dynamin Vps1 (Figure 4) [60].

## 5. The Recycling Pathway

After endocytic internalization, proteins are transported to endosomes and then further delivered to the vacuole to be degraded as described above, but several proteins are recycled back to the PM after internalization. Thus they are not targeted to the vacuole, but undergo retrograde transport from the endosomes to the TGN where they integrate the secretory pathway.

The recycling pathway (RCY) plays an economical role in the cells as it allows the reuse of proteins such as receptors (Figure 2). Indeed, in many instances their recycling participates in signal attenuation by removing the agonist from the medium. In general the ligands are detached from their receptors at the level of early endosomes, follow the endocytosis pathway, and end up in the vacuolar lumen, whereas the receptors can return to the PM and bind to another ligand.

### 5.1. Recycling of the SNARE Snc1

Important families of proteins that undergo recycling are membrane bound SNARE (Soluble *N*-ethylmaleimide-sensitive-factor Attachment protein REceptor) proteins required for vesicles/compartments fusion. In yeast *S. cerevisiae*, the best studied cargo of the recycling pathway is the v-SNARE Snc1. Snc1 is mainly localized at the PM at the sites of secretion as the emerging bud. Snc1 mediates the fusion between secretory vesicles and the plasma membrane, is then internalized and undergoes recycling from the early endosomes to the TGN from where it is re-secreted to the plasma membrane. The Snc1 recycling requires the GTPases Ypt31/32 and their effector protein Rcy1 [61,62].

### 5.2. Recycling of the Chitin Synthase Chs1 and Chs3

Chitin is one of the most abundant components of the yeast cell wall and its synthesis depends on the chitin synthase enzymes Chs1, Chs2, and Chs3. Chs3 (the main chitin synthase) and Chs1 are stored in intracellular compartments termed chitosomes and transported to the PM when required [63,64]. This targeting to the PM is dependent on the recycling pathway; when Chs3 is not required at the PM anymore it is internalized via endocytosis and targeted back to the chitosome.

The origin and composition of this chitosome is still under debate, and it is suggested to be a product of endocytosis that is also linked to exocytosis. Indeed, the retention of Chs3 into the internal reservoir requires the Chs6 protein, the AP-1 adaptor complex, and its binding proteins the epsins Ent3 and Ent5 [65,66]. The epsins Ent3/Ent5 are required for TGN to endosomal trafficking and for late endosomal/MVB sorting [40,41]. Here, their proposed function is to mediate the transport of Chs3 from the TGN to the chitosomes (that is, an endosome derived compartment), and thus in their absence the TGN localized Chs3 is secreted to the PM [66]. Chs1 is subjected to the same regulation as Chs3 but does not depend on the same effector, for example, its chitosome targeting is not Chs6-dependent [63].

## 6. New Trafficking Pathways

Since the first genetic screen done in the early 1980s in the laboratory of Randy Schekman to isolate membrane trafficking genes, research has mainly focused on vesicular transport between organelles. In recent years it has come to light that the contact sites between organelles and intracellular compartments are also playing an important role for the cellular homeostasis. Indeed, contact sites between mitochondria and the ER observed in the 1950s by Copeland and Dalton were shown to be lipid exchange sites, thus solving the small conundrum, that is, lipid synthesis occurring at the ER membrane but requiring some mitochondrial modifications [67,68]. These exchange sites rely on a protein complex called ERMES (endoplasmic reticulum-mitochondrial encounter structure) identified in yeast and shown to tether one membrane to the other; this ERMES structure is also found in other eukaryotes [69]. Contact sites between the ER and the Golgi, chloroplasts, vacuoles/lysosomes, endosomes and the plasma membrane were also identified [70,71]. The function of these contacts sites is not well known at the moment. Other such interactions have been found between the nucleus and the vacuole called Nucleus-Vacuole Junctions (NVJs) and allowing piecemeal autophagy of the nucleus and its degradation in the vacuole [72]. The most recent inter-organellar contact site, described by both Maya Schuldiner's and Christian Ungermann's laboratories, is located between the vacuole and the mitochondria and is called vCLAMP (vacuole and mitochondria patch) [73,74]. Although the function of this contact site is not clear yet, it relies on the interaction between the Ypt7 Rab-GTPase and the Vps39 HOPS complex subunit [74]. Interestingly, deletion of either the ERMES or the vCLAMP complex causes the expansion of the other system to complement for the loss-of-function, and the deletion of both is lethal; moreover overproduction of vCLAMP can also suppress ERMES deletion phenotypes [73,74]. These recent studies highlight the role of interorganellar contact sites in membrane trafficking.

## 7. Conclusions

The yeast *S. cerevisiae* has been influential in identifying protein complexes required for the transport of proteins from one organelle to the other. Although vesicular transport is now well understood, yeast is again instrumental in the understanding a new type of trafficking between organelle by direct membrane-membrane contact sites [75]. Since its vesicular trafficking is well characterized, yeast is also becoming a very amenable mode to study human diseases for which the cause has been mutation in genes encoding membrane trafficking proteins [76]. These new developments show how this amenable model will be still relevant in cell biology research for the foreseeable future.

## Acknowledgments

This work was supported by the CNRS (ATIP-CNRS 05-00932 and ATIP-Plus 2008-3098 to S. Friant), the Fondation Recherche Médicale (FRM INE20051105238 and FRM-Comité Alsace 2006CX67-1 to S. Friant and FRM Postdoctoral fellowship to J-O. De Craene), the Association pour la Recherche sur le Cancer (ARC JR/MLD/MDV-CR306/7901 to S. Friant), Agence Nationale de la Recherche (ANR-07-BLAN-0065 and ANR-13-BSV2-0004), the Association Française contre les Myopathies (AFM-SB/CP/2013-0133/16551 grant to S. Friant and fellowship to D.L. Bertazzi) and the Fonds National de la Recherche (FNR, Luxembourg) doctoral fellowship to S. Feyder.

## Author Contributions

All authors participated in the writing of the paper.

## Conflicts of Interest

The authors declare no conflict of interest.